\documentclass[reprint,amsmath,amssymb,aps,prl,showpacs]{revtex4-1}

\usepackage{graphicx}
\usepackage{dcolumn}
\usepackage{bm}
\usepackage{hyperref}

\begin{document}

\title{High-field magnetoresistance revealing scattering mechanisms in graphene}

\author{Andreas Uppstu}
\author{Ari Harju}

\affiliation{
COMP Centre of Excellence and Helsinki Institute of Physics, Department of Applied Physics, Aalto University School of Science, FI-02150 Espoo, Finland
}

\date{\today}

\begin{abstract}
We show that the type of charge carrier scattering significantly affects the high-field magnetoresistance of graphene nanoribbons. This effect has potential to be used in identifying the scattering mechanisms in graphene. The results also provide an explanation for the experimentally found, intriguing differences in the behavior of the magnetoresistance of graphene Hall bars placed on different substrates.  Additionally, our simulations indicate that the peaks in the longitudinal resistance tend to become pinned to fractionally quantized values, as different transport modes have very different scattering properties.
\end{abstract}

\maketitle
Despite lots of both experimental and theoretical effort, the issue of charge carrier scattering in graphene has not been completely settled \cite{RevModPhys.83.407, Adam, PhysRevLett.107.225501, PhysRevLett.98.186806, PhysRevB.79.245423, PhysRevB.76.205423, PhysRevB.76.205423, PhysRevB.83.165402}. On ${\mathrm{SiO}}_{2}$ substrates, scattering by long-ranged Coulomb impurities has been proposed to be a major factor contributing to the conductivity near the Dirac points \cite{RevModPhys.83.407}. Specifically, the measured minimum conductivity in samples with a large width to height ratio is explained by such scattering \cite{Adam, RevModPhys.83.407}. Also short-ranged scattering is thought to occur in graphene, caused by resonant scatterers, like adsorbed hydrocarbons or hydrogen atoms \cite{Ni} or, in the case of narrow graphene nanoribbons, edge roughness \cite{PhysRevB.79.075407}. As charge carrier scattering significantly affects the performance of graphene-based devices, knowledge of the scattering mechanisms is essential to realize the full potential of the material.

A recent study found that the behavior of the high-field magnetoresistance of graphene on a ${\mathrm{SrTiO}}_{3}$ substrate is clearly different from that on a ${\mathrm{SiO}}_{2}$ substrate \cite{PhysRevLett.107.225501}. Specifically, the magnetoresistance peak corresponding to the zeroth Landau level (LL) decreases as a function of decreasing temperature on ${\mathrm{SrTiO}}_{3}$, whereas previous studies have shown that it increases with decreasing temperature on ${\mathrm{SiO}}_{2}$ \cite{PhysRevLett.100.206801, PhysRevB.80.201403, PhysRevLett.105.046804, PhysRevLett.98.196806}. Although the reason for this qualitative difference has not yet been found, the results lead to strong argumentation about the dominant scattering mechanism in graphene \cite{PhysRevLett.107.225501, DasSarma20121795}.

In this Letter, we simulate high-field magnetotransport in graphene nanoribbons (GNRs) and present results that agree qualitatively with the measurements. As our calculations will show, the magnetoresistance of the zeroth LL depends  strongly on the type of the disorder, implying that high-field magnetoresistance measurements may be used to probe the scattering mechanisms in graphene. We also find that the magnetoresistance peaks at the other LLs tend to become pinned to fractional values of the resistance quantum, as the localization lengths of different transport modes differ significantly.

Our results are based on an orthogonal tight-binding model, with hoppings taken up to third nearest neighbors \cite{PhysRevB.81.245402}. The model system is a two-terminal device with semi-infinite leads. Compared with a Hall bar, a two-terminal model is computationally less expensive and gives more fundamental information about the scatterers, as the explicit design of the voltage probes is not taken into account. It has to be noted that models that take electron-electron interactions into account predict spin-polarized states \cite{PhysRevB.80.235417}, while our computations apply to systems with degenerate spins.

The transmission is computed using standard Green's function based methods and the conductance is acquired from the Landauer formula. We simulate charged impurities with Gaussian potentials $V=A \exp [-(\mathbf{r}-\mathbf{r_0})^2/2\sigma^2]$, with the parameter $A$ selected to be $+0.1$ eV for half of the impurities and $-0.1$ eV for the rest, in order to prevent doping. The amplitude of the potentials roughly corresponds to the measured amplitude of the potential variation on the surface of ${\mathrm{SiO}}_{2}$ \cite{Martin}. As the dielectric constant of ${\mathrm{SrTiO}}_{3}$ increases from 200-300 to 5000 when the temperature is lowered from room temperature to liquid helium temperature \cite{PhysRev.174.613, PhysRevB.19.3593}, we select the range $\sigma$ of the potentials as a variational parameter. Shorter ranged potentials represent charged impurities on ${\mathrm{SrTiO}}_{3}$ at low temperatures, and longer ranged those at higher temperatures or on ${\mathrm{SiO}}_{2}$.

\begin{figure}
\vspace{0.3cm}
~~\includegraphics[width=\columnwidth]{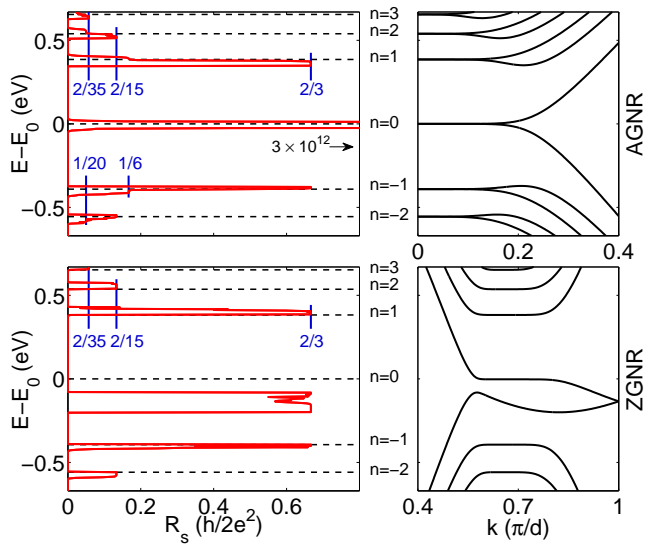}
\caption{(Color online) Scatterer resistance $R_s$ for GNRs with long-range Gaussian potentials at $B=200$ T with respect to the location of the zeroth LL $E_0$ (left), and the band structures of pristine ribbons (right). The upper row corresponds to a 15 nm wide AGNR, and the lower row to an equally wide ZGNR. The dashed lines in the $R_s$ plots indicate the positions of the LLs in the pristine ribbons. Resistance values predicted by Eq.~(\ref{q1}) for the maximum values of the peaks and Eq.~(\ref{q2}) for the shoulders of the peaks of the AGNR are also shown.}
\label{f1}
\end{figure}

We start by comparing the scatterer resistances $R_s$ of a 500 nm long and 15 nm wide armchair-edged GNR (AGNR) and a zigzag-edged GNR (ZGNR) in left hand side panel of Fig.~\ref{f1}. We acquire $R_s$ from the total resistance $R_\text{tot}$ by using the expression $R_s=R_\text{tot}-1/M$, where the last term is the contact resistance with $M$ being the number of transport modes. All resistances are given in units of $h/2e^2$. Each ribbon contains Gaussian impurity potentials at a concentration of $1.3 \times 10^{11}$ cm$^{-2}$, which is of the same order of magnitude as an estimate based on scanning tunneling microscopy of graphene on ${\mathrm{SiO}}_{2}$ \cite{Crommie}. The range of the potentials is set to 15 times the C--C distance $a$, and the strength of the perpendicular magnetic field is 200 T. With wider ribbons, weaker magnetic fields are required to produce similar results. The band structures of pristine ribbons are shown in the right hand panel. If the spin degeneracy is not taken into account, the band structure of the AGNR contains a single valley with doubly degenerate LLs, whereas the ZGNR exhibits two valleys each with nondegenerate LLs. In AGNRs, the degeneracy of the LLs is broken near the edges of the ribbon \cite{PhysRevLett.107.086601}. It can be seen that $R_s$ is zero between the LLs, as the physical separation of the opposite edge channels, corresponding to charge carriers moving in opposite directions, prevents backscattering. At the LLs the charge carrier density extends over the central part of the ribbon, and peaks in the scatterer resistance are formed.

A study of the resistance peaks reveals fractional quantization \cite{Datta, Beenakker}, due to significantly different scattering properties of the transport modes. While modest disorder causes the inner channels to become almost completely localized, the outer channels remain almost completely conducting. The total number of channels that correspond to charge carriers moving in a specific direction (left or right) at the $n$th LL in a GNR is given by  $2|n|+1$, and in the case of the ZGNR, the innermost channels in both valleys are localized. Thus, for LLs with $|n|>1$, we obtain $R_s=1/(2|n|-1)-1/(2|n|+1)$, which simplifies to
\begin{equation}
R_s=\frac{2}{4n^2-1}, ~~~~~~~ |n| > 0.
\label{q1}
\end{equation}
Additionally, a resistance peak of 2/3 is found just below the zeroth LL, corresponding to two of the three channels being localized. In the case of the AGNR, on the other hand, all channels at the zeroth LL are localized, resulting in a very high peak in the resistance. Furthermore, when $|n| > 0$, two kinds of behavior are observed. On the inside (closer to half filling) of each LL, the two innermost channels are localized, yielding the same quantization expression as for the ZGNRs, i.e. Eq.~(\ref{q1}). However, on the outside of the LLs (farther away from half filling), only the innermost channels are localized, leading to the expression
\begin{equation}
R_s=\frac{1}{4n^2+2|n|},~~~~~~~ |n| > 0.
\label{q2}
\end{equation}
Because of this, in an AGNR, the peaks in $R_s$ tend to exhibit steps or shoulders when $|n| > 0$.

A comparison with experimental data given in Ref.~\cite{Kim} shows that the measured peaks in the longitudinal resistance of a multiterminal Hall bar of micrometer size differ roughly by 30~\% from the values given by Eq.~(\ref{q1}). Specifically, the measured resistance at the $n=1$ to $n=3$ LLs is lower than the prediction by Eq.~(\ref{q1}), which would correspond to a situation where the inner channels are not yet fully localized. Additionally, the two-terminal model does not take into account effects caused by disordered contacts.  

\begin{figure}
\vspace{0.3cm}
~~\includegraphics[width=\columnwidth]{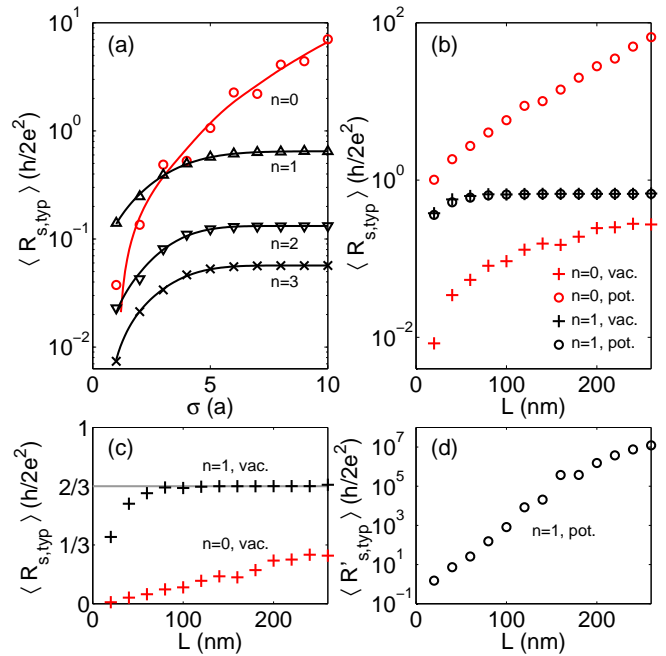}
\caption{(Color online) Statistical results showing the heights of peaks in the energy-averaged typical scatterer resistance, corresponding to the lowest Landau levels, in a magnetic field of 200 T. (a) Peak heights as a function of the range of the potentials in a system with a constant concentration of Gaussian impurities. The lines are to guide the eye. (b) Peak heights as a function of ribbon length for both Gaussian potentials and vacancies. The magnetic field is 200 T. (c) The results for the vacancies shown on a linear scale, highlighting the linear growth of the $n=0$ peak. (d) Growth of $\langle R'_\text{s,typ} \rangle$ for monvacancies when $n=1$.}
\label{f2}
\end{figure}

We next study how the range of the potentials affects the heights of the peaks in $R_s$. We define the energy-averaged typical scatterer resistance as $\langle R_{s,\text{typ}} \rangle \equiv 1/\langle \exp (\overline{\log (T_\text{tot})}) \rangle -1/ \langle M \rangle$~\cite{PhysRevB.22.3519}, where $T$ is the transmission function, the overlined quantity is ensemble averaged over random impurity configurations and the brackets stand for averaging over a finite energy window, $\Delta E$, corresponding to a finite bias voltage of $V=\Delta E /e$. Here we have set $\Delta E$ to 1 meV. A statistical study of the peak heights in a 15 nm wide AGNR is presented in Fig.~\ref{f2}, where (a) shows how $\langle R_{s,\text{typ}} \rangle$ evolves in a 15 nm wide and 100 nm long AGNR as the range of the impurity potentials is gradually changed, keeping the concentration constant at 6.7 $\times 10^{11}$ cm$^{-2}$. Each data point is based on an ensemble of 50 different realizations of impurity locations and transmission averaging over 20 energy points within the bias window. With short-ranged potentials, the central peak, corresponding to the $n=0$ LL, is shorter than the neighboring ones, but it increases drastically as $\sigma$ is made larger, as the charge carriers become localized. The peaks corresponding to other LLs increase only toward the quantized values, as only the inner modes become localized. The odd behavior of the $n=0$ peak may be attributed to the chiral symmetry of the graphene lattice. According to Ref.~\cite{PhysRevLett.103.156804}, in the presence of disorder that preserves the chiral symmetry, the $|n|>0$ LLs become broadened, whereas the zeroth LL remains delta-function-like. In our simulations, modest concentrations of short-ranged scatterers or relatively short-ranged Gaussian potentials, that cause a clear broadening of the resistance peaks at the $|n|>0$ LLs, do not lead to a significant broadening the zeroth LL, if compared with the experimentally relevant energy scale set by a small bias voltage. This indicates that modest disorder affects the chiral symmetry of graphene only weakly, whereas stronger potential disorder, simulated by longer-ranged Gaussian impurities, leads to a broadening of all LLs.

Fig.~\ref{f2} (b) shows the energy-averaged typical scatterer resistances at $n=0$ and $n=1$ as functions of ribbon length, for a constant concentration (6.7 $\times 10^{11}$ cm$^{-2}$)  of impurity potentials with range $\sigma=15a$, as well as for a constant concentration (1.7 $\times 10^{12}$ cm$^{-2}$) of  monovacancies. In the presence of the potentials, the resistance at $n=0$ grows exponentially with length due to strong localization, whereas the resistance at $n= 1$ approaches the fractionally quantized value of 2/3, given by Eq.~(\ref{q1}). On the other hand, when the disorder is formed by vacancies, the $n=0$ LL has clearly smaller resistance than the $n= 1$ LL. As shown in the linear plot of Fig.~\ref{f2} (c), the $n=0$ peak grows linearly, whereas the $n=1$ peak is already close to the fractionally quantized value, indicating strong localization of the inner modes. Although not shown in the figure, strong enough disorder may cause scattering also in the outer channels, and thus an unpinning of the fractionally quantized resistance.

\begin{figure}
\vspace{0.3cm}
\includegraphics[width=\columnwidth]{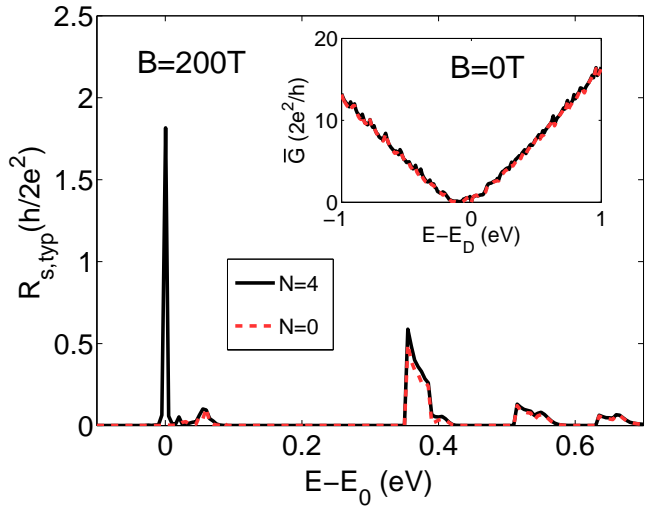}
\caption{(Color online) Typical scatterer resistance $R_{s,\text{typ}}$ of a 120-AGNR with vacancies but no long-range disorder (dashed red/gray), and in the presence of both vacancies and long-ranged disorder formed by $N=4$ Gaussian potentials (black). The peak at the zeroth LL has been averaged over an energy window of 1 meV. The inset shows the corresponding average conductances with respect to the Dirac point $E_D$ at $B=0$ T.}
\label{f3}
\end{figure}

The energy-averaged resistance corresponds to a localization length $\xi$~\footnote{We have defined $\xi$ so that the resistance grows as $\exp(L/\xi)$ in the localized regime} of roughly 60 nm at the zeroth peak for potentials with $\sigma=15a$. This can be turned into a scattering cross section $\Sigma$ of 2.5 nm by using the expression $\Sigma=WLM/N\xi$ \cite{PhysRevB.85.041401, PhysRevB.81.125307}, valid for systems belonging to the unitary transport ensemble \cite{RevModPhys.69.731}, where $N$ is the number of impurity potentials and $W$ and $L$ the width and length of the ribbon, respectively. The scattering cross section is strictly valid only for a 15 nm wide AGNR, as the arbitrarily located potentials extend partly outside the ribbon edges. When $|n|>0$, we approximate that only $M'$ of the totally $M$ modes exhibit scattering, and express the total resistance in the localized regime as $R_\text{tot}^{-1}=M'\exp(-L/\xi')+M-M'$, where $\xi'$ is the localization length corresponding to the $M'$ modes only. Fig.~\ref{f3} (c) shows  $R'\equiv \exp(L/\xi')/M'$ for the lowest LLs with $\sigma=15a$, indicating a clearly localized behavior, with $\xi'=$ 15 nm and $\Sigma'=$ 20 nm. The inner modes at $n=1$ show a larger scattering cross section than the modes at $n=0$, but the almost perfectly conducting outer modes limit the resistance to a fractionally quantized value. As the localization lengths of the inner modes are very short, one has to be careful when interpreting resistivities given by the Ohmic formula $\rho=RL/W$ \cite{0295-5075-10-5-017}.

If the Gaussian potentials are replaced by monovacancies, $\Sigma'$ for the $n=1$ LL decreases to 8 nm, while the resistance peak at $n=0$ grows linearly as $R_s=N \Sigma /W$, yielding $\Sigma = $ 0.065 nm. This indicates a qualitative difference between long- and short-ranged impurities, as the latter scatter charge carriers only marginally at the zeroth LL. When the range of the potentials is short enough, the $n=0$ peak is much smaller than the $n= \pm 1$ peaks, but once the range increases, the $n=0$ peak grows without limits, whereas the growth of other peaks is limited by the presence of modes that exhibit only very limited scattering.

Finally, we compare the effects of impurity disorder in high magnetic field with those in zero magnetic field, using a test system that also has short-ranged disorder. Fig.~\ref{f3} shows $\langle R_{s,\text{typ}} \rangle$ computed from an ensemble of 100 AGNRs of length 200 nm and width 15 nm, both with and without long-ranged impurity potentials, and with vacancies at a concentration of 6.7 $\times 10^{11}$ cm$^{-2}$. With no impurity potentials, the peak corresponding to the zeroth LL is negligible. If bulk vacancies are replaced by edge disorder, similar results are acquired (not shown). When four impurity potentials (two positive and two negative), with range $\sigma=8a$, are added to the system, a high resistance peak is formed at $n=0$. This contrasts clearly against the behavior at zero magnetic field, shown in the inset. The ensemble averaged conductance of an AGNR at $B=0$ T shows that the addition of impurity potentials does not produce any significant change in the conductance of a GNR in the absence of a magnetic field.

Our results are valid for phase coherent systems. When the ribbon is longer than the phase coherence length, localization effects will increase with lowering temperature, until the ribbon becomes fully phase coherent. Unless already pinned to quantized values, the heights of the peaks in $R_s$ are expected to increase when the temperature is lowered, if the system is
longer than the corresponding localization lengths. Importantly, the $n=0$ peak of an AGNR does not become pinned, which may correspond to the behavior observed in Hall bar experiments on ${\mathrm{SiO}}_{2}$ \cite{PhysRevLett.100.206801, PhysRevB.80.201403, PhysRevLett.105.046804, PhysRevLett.98.196806}. On the other hand, in the recent experiment on ${\mathrm{SrTiO}}_{3}$~\cite{PhysRevLett.107.225501}, a decrease of the temperature and thus an increase of the substrate dielectric constant produced a decreasing peak at the $n=0$ LL, whereas the heights of the peaks at the $|n|>0$ LLs increased. According to our simulations, this is explained by a reduced range of the impurity potentials. At zero magnetic field, no change in conductivity was found with decreasing temperature \cite{PhysRevLett.107.225501}, but our simulations indicate that this does not necessarily imply that charged impurities are absent. As the inset in Fig.~\ref{f3} shows, at zero magnetic field scattering by potential disorder may be completely masked by scattering by short-ranged disorder. On the other hand, in two-dimensional graphene, the phenomenon of minimum conductivity will prevent the observation of localization effects \cite{RevModPhys.83.407, PhysRevB.79.245423, Adam}.

To conclude, in graphene nanoribbons, the magnetoresistance at the lowest Landau level depends strongly on the type of the disorder. At higher Landau levels, and at the zeroth Landau level in zigzag edged ribbons, the magnetoresistance may become pinned to fractionally quantized values due to significantly different conductance properties of the inner and outer transport modes. In armchair-edged ribbons, the resistance peaks at higher Landau levels tend to exhibit shoulders. According to our results, a decrease in the range of impurity potentials will cause a significant decrease in the measured magnetoresistance at the zeroth Landau level, which is in line with recent experimental results on a ${\mathrm{SrTiO}}_{3}$ substrate.

\begin{acknowledgments}
 This research has been supported by the Academy of Finland through its Centres of Excellence Program (project no. 251748). We thank Mikko Ervasti, Ilja Makkonen and Antti-Pekka Jauho for commenting on the manuscript.
\end{acknowledgments}

\end{document}